\pdfoutput=1
\documentclass[twocolumn,prd,superscriptaddress,preprintnumbers,nofootinbib]{revtex4}
\usepackage{graphicx}
\usepackage{epsfig}
\usepackage{lipsum}
\usepackage{enumitem}
\usepackage{bm}
\usepackage{latexsym,amssymb,amsmath,amsfonts,amssymb,txfonts,pxfonts,wasysym,float}
\usepackage{color}

\newcommand{\beq}[1]{\begin{equation}\label{#1}}
\newcommand{\eeq}{\end{equation}}
\newcommand{\bea}[1]{\begin{eqnarray} \label{#1}}
\newcommand{\eea}{\end{eqnarray}}
\newcommand{\ba}{\begin{array}}
\newcommand{\ea}{\end{array}}

\def\be{\begin{equation}}
\def\ee{\end{equation}}
\def\gs{\mathrel{
   \rlap{\raise 0.511ex \hbox{$>$}}{\lower 0.511ex \hbox{$\sim$}}}}
\def\ls{\mathrel{
   \rlap{\raise 0.511ex \hbox{$<$}}{\lower 0.511ex \hbox{$\sim$}}}}

\newcommand{\postscript}[2]{\setlength{\epsfxsize}{#2\hsize}
   \centerline{\epsfbox{#1}}}

\newcommand{\comment}[1]{}

\usepackage[usenames,dvipsnames]{xcolor}
\definecolor{orange}{cmyk}{0,0.5,1,0}
\definecolor{rossoCP3}{cmyk}{0,.88,.77,.40}
\definecolor{graa}{rgb}{0.8,0.8,0.8}
\definecolor{blaa}{rgb}{0.2,0.2,0.6}

\begin{document}

\title{\color{rossoCP3}{Hunting for super-heavy dark matter with the highest-energy cosmic
    rays}}


\author{Esteban Alcantara}

\affiliation{Department of Physics and Astronomy,  Lehman College, City University of
  New York, NY 10468, USA
}

\author{Luis A. Anchordoqui}

\affiliation{Department of Physics and Astronomy,  Lehman College, City University of
  New York, NY 10468, USA
}

\affiliation{Department of Physics,
 Graduate Center, City University
  of New York,  NY 10016, USA
}

\affiliation{Department of Astrophysics,
 American Museum of Natural History, NY
 10024, USA
}

\author{Jorge F. Soriano}

\affiliation{Department of Physics and Astronomy,  Lehman College, City University of
  New York, NY 10468, USA
}

\begin{abstract}
  \vskip 2mm \noindent In 15 years of data taking the Pierre Auger
  Observatory has observed no events beyond $10^{11.3}~{\rm
    GeV}$. This null result translates into an upper bound on the flux
  of ultrahigh-energy cosmic rays implying
  $J (> 10^{11.3}~{\rm GeV}) < 3.6 \times 10^{-5}~{\rm km}^{-2} {\rm
    sr}^{-1} {\rm yr}^{-1}$, at the 90\%C.L. We interpret this bound
  as a constraint on extreme-energy photons originating in the decay
  super-heavy dark matter (SHDM) particles clustered in the Galactic
  halo. Armed with this constraint we derive the strongest lower limit
  on the lifetime of hadronically decaying SHDM particles with masses
  in the range, $10^{14} 􏰹\alt M_X/{\rm GeV} \alt 10^{16}$. We also
  explore the capability of future NASA's POEMMA mission to search for
  SHDM signals.
\end{abstract}
\maketitle

\section{Introduction}

For the time being, a sovereign objective of the particle physics
program is to ascertain the connection between dark matter (DM) and
the Standard Model (SM). Existing data constrain the majority of DM to
be non-baryonic, cold or warm, and stable or
long-lived~\cite{Bertone:2004pz}. There are many ways to accommodate
these constraints and so feasible DM candidates with a very large
range of masses and interaction strengths have been
proposed~\cite{Feng:2010gw}.

For many decades, the favored models characterized the DM as a relic
density of weakly interacting massive particles
(WIMPs)~~\cite{Lee:1977ua,Vysotsky:1977pe,Goldberg:1983nd,Steigman:1984ac}.\footnote{For a precise calculation of the  WIMP
 relic abundance,  see~\cite{Gondolo:1990dk,Steigman:2012nb};  partial wave unitarity dictates an upper bound on the WIMP mass
$\leq 110~{\rm TeV}$~\cite{Griest:1989wd,Blum:2014dca}.}  However, LHC
experiments have run extensive physics searches for WIMP signals which
have returned only null
results~\cite{Buchmueller:2017qhf,Penning:2017tmb}. In addition, a
broad WIMP search program has been developed with direct and indirect
detection methods, which so far have given unsatisfactory
answers~\cite{Undagoitia:2015gya,Aprile:2017iyp,Akerib:2016vxi,Cui:2017nnn,Amole:2017dex,Aartsen:2017ulx,
  Albert:2016emp,Ahnen:2016qkx,Abdallah:2018qtu,Albert:2017vtb,Abeysekara:2017jxs,Ackermann:2012rg}. Despite
the fact that a complete exploration of the WIMP parameter space remains
the highest priority of the DM community, there is now a strong
motivation to explore alternatives to the WIMP paradigm.

Among the well-motivated ideas for what DM could be, the WIMPzilla
hypothesis postulates that DM is made of gravitationally produced
(non-thermal relic) superweakly-interacting supermassive
$X$-particles~\cite{Chung:1998zb,Kuzmin:1998uv,Kuzmin:1998kk,Kolb:1998ki,Chung:1999ve,Kuzmin:1999zk,Chung:2001cb,Kolb:2007vd,Kannike:2016jfs}. As
a matter of fact, the gravitational production of superheavy dark
matter (SHDM) at the end of inflation may be taken as the only
experimentally verified DM production mechanism, because the observed
cosmic microwave background (CMB) fluctuations have precisely the same
origin. At the end of inflation a fraction of fluctuations are not
stretched beyond the horizon but remain as $X$-particles because the
inflation slows down. The weakness of the gravitational interaction
naturally explains the tiny initial abundance of WIMPzillas. Indeed,
for such an abundance to be cosmologically relevant today, the
$X$-particles must be supermassive.

On an entirely separate though somewhat related note, the surprising
absence of any signals of new physics at the LHC
experiments~\cite{Rappoccio:2018qxp} seems to indicate that nature
does not too much care about our notion of naturalness. Indeed the
required fine-tuning of SM fundamental parameters to accommodate the
15 orders of magnitude between the electroweak and the Planck scales
may soon become a reality. Of course, the only reason one may try to
incorporate such a shocking idea is that the existence of life may
actually be contingent on this wicked
conspiracy~\cite{Weinberg:1987dv}. Namely, the weak and QCD scales
come about just very close to one another, so that a plethora of atoms
can exist to exchange energy over extremely long timescales,
assembling the building blocks for life and durable habitats where it
can
thrive~\cite{Agrawal:1998xa,Susskind:2003kw,Feldstein:2006ce,Donoghue:2009me}.\footnote{Investigations
  in String Theory have applied a statistical approach to the enormous
  ``landscape'' of vacua present in the
  theory~\cite{Susskind:2003kw}. Remarkably, these huge number of
  metastable vacua, ${\cal O} (10^{500})$, can also accommodate the
  more severe fine-tuning required to characterize the SM with a small
  cosmological constant~\cite{Douglas:2003um,Ashok:2003gk}.} An
additional, though not so severe, anthropic argument applies to the
abundance of DM, which cannot be too much larger or smaller than what
is
observed~\cite{Wilczek:2004cr,Hellerman:2005yi,Tegmark:2005dy,Freivogel:2008qc}. This
is because DM plays a critical role in structure formation. Note that
since DM is only subject to the force of gravity, the gravitational
Jeans instability which allows compact structures to form is not
opposed by any force, such as radiation pressure. As a result, DM
begins to collapse into a complex network of DM halos well before
baryonic matter, which is impeded by pressure forces. Without DM, the
epoch of galaxy formation would occur substantially later in the
universe than is observed, and consequently the galaxies needed for
our existence would not have formed in time. However, it is only the
DM abundance and not any other details of the dark sector which is
critical for life to exist. Therefore, it is quite reasonable to
expect that the DM sector would not be as fine tune as the visible SM
sector.  In other words, even if we are prepared to advocate the
anthropic argument to accommodate the unnaturalness of the weak scale,
we would expect the DM particle spectrum to
be as natural as possible, i.e. near the Planck scale that is the natural
ultraviolet cutoff scale. For the
most part, the WIMPzilla could then be a natural DM candidate and
perhaps as well-motivated as the WIMP paradigm.

Furthermore, precision CMB measurements enable a direct experimental
test of the WIMPzilla hypothesis.  This is because the production of
SHDM during inflation gives rise to isocurvature perturbations that
become sources of gravitational potential energy contributing to the
tensor power spectrum of the CMB~\cite{Chung:2004nh}. This implies a
detectable primordial tensor-to-scalar ratio $r$ in the CMB power
spectrum. The combined (Planck satellite~\cite{Ade:2015lrj} together
with BICEP2 and the Keck array~\cite{Array:2015xqh}) 95\% C.L.  upper
bound, $r < 0.07$, already constrains the $X$-particle mass to be
$M_X \alt 10^{17}~{\rm GeV}$ in the limit of instantaneous
reheating~\cite{Garny:2015sjg}. For slightly less efficient reheating,
this upper limit strengthens to $M_X \alt 10^{16}~{\rm GeV}$. 

Note also that while the WIMPzilla must be stable over cosmological timescales,
instanton decays induced by operators involving both the hidden sector
and the SM sector may give rise to observable signals in the spectrum
of ultrahigh-energy cosmic rays
(UHECRs)~\cite{Kuzmin:1997jua,Berezinsky:1997hy}. More concretely, the
spectrum from WIMPzilla decay is expected to be dominated by photons
and neutrinos because of a more effective production of pions than
nucleons in the QCD cascades. Since the photons would not be
attenuated owing to their proximity, they become the prime signal
because it is easier to detect photons than neutrinos. In this article 
we use the most recent UHECR data to derive the strongest lower limit
on the lifetime of hadronically decaying WIMPzillas. We also investigate
the prospects for next generation UHECR experiments to search for
SHDM signals.

\section{New limit on the lifetime of SHDM}

The Pierre Auger Observatory has collected an exposure
${\cal E} = 67,000~{\rm km^2 \, sr \, yr}$ without observation of any events
with energy $E_0 > 10^{11.3}~{\rm GeV}$~\cite{Aab:2017njo}. This null result sets a
generic upper limit on the integrated flux of UHECRs; namely,
\begin{eqnarray}
  J (> E_0) = \int_{E_0}^\infty  J (E) \, dE &
<& 2.44/{\cal E}
                                                 \nonumber \\
  & < & 3.6 \times 10^{-5}~{\rm
    km}^{-2} {\rm sr}^{-1} {\rm yr}^{-1} \, ,
\label{gammabound}
\end{eqnarray}
at the 90\% C.L.; the limit is a factor of 1.266 less restrictive at the 95\% C.L.~\cite{Feldman:1997qc}. When interpreted as a bound on
extreme-energy photons and compared with existing bounds~\cite{Aab:2016agp,Abbasi:2018ywn}, this limit
is more restrictive by about an order of magnitude, but at a slighter higher
energy. Consequently, the all-particle limit of (\ref{gammabound}) could provide a better weapon to constrain
WIMPzilla decay.

\begin{figure}[tb] 
    \postscript{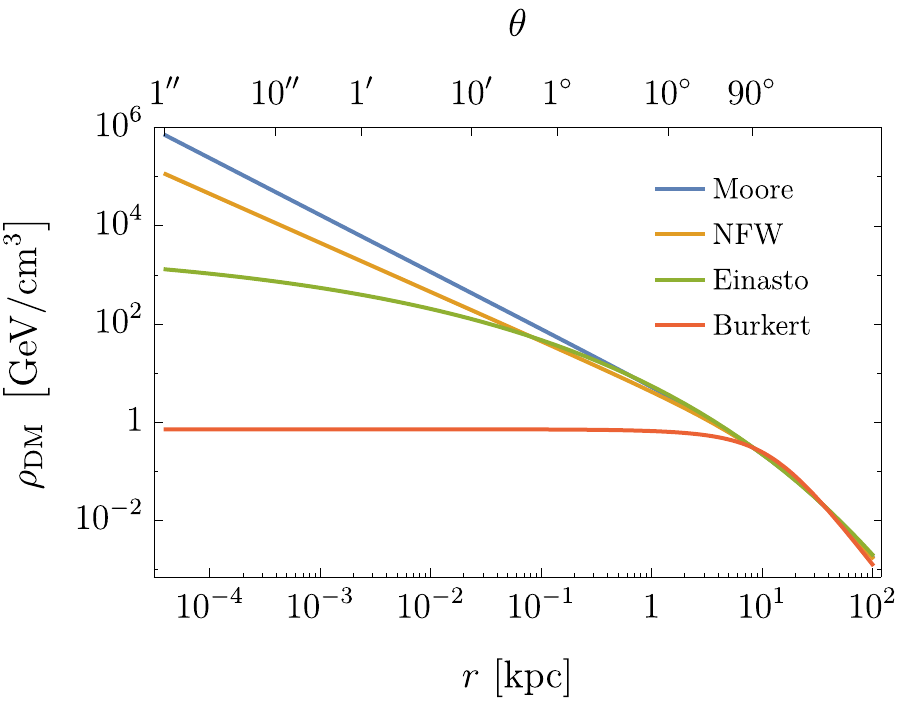}{0.99} 
\caption{DM halo mass profiles. The upper horizontal axis
  shows the variation of the angle between the line of sight and the axis
defined by Earth and the Galactic center. \label{fig:1}}
\end{figure}

\begin{figure}[tb] 
    \postscript{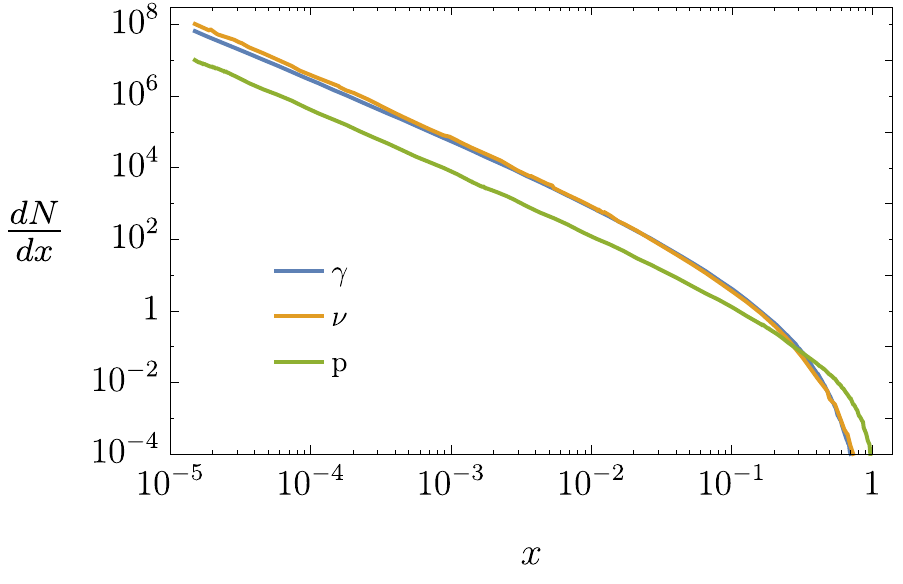}{0.99} 
    \caption{Spectra of photons, protons, and neutrinos ($\nu +\bar
      \nu$) from $X$ particle decay as a function of the dimensionless
      variable $x = 2E/M_X$. We have taken $M_X = 10^{16}~{\rm GeV}$. \label{fig:2}}
\end{figure}

To estimate the photon flux from WIMPzilla decay we need to evaluate
two separate contributions: the astrophysical factor and the particle
physics factor.
\begin{itemize}[noitemsep,topsep=0pt]
\item The
astrophysical factor is determined by the distribution of DM
particles in the Galaxy. The DM density of X-particles is a function
of the distance $r$ from the Galactic Center and  is usually described by a smooth profile function
\begin{equation} 
\rho_{\rm DM} (r)=\frac{\rho_s}{\left[1 - \beta + (r/r_s)^\alpha\right] (1+r/r_s)^{3-\alpha}} \,,
\label{densityX}
\end{equation}
where $\rho_s$ and $r_s$ are respectively the scale density and scale
radius. The traditional benchmark choice, motivated by N-body simulations, is the
Navarro-Frenk-White (NFW) profile, in which $\alpha
=1$, $\beta = 1$, and $r_s = 24.42~{\rm
  kpc}$~\cite{Navarro:1996gj}. The latest 
numerical simulations, however, seem to favor the Einasto profile,
\begin{equation}
\rho_{\rm DM} (r) = \rho_s \exp \left\{- \frac{2}{0.17}
\left[\left(\frac{r}{r_s}\right)^{0.17}  - 1 \right]\right\} \,, 
\end{equation}
which does not converge to a power law at the Galactic Center and
becomes more chubby than NFW at kpc scales, and where
$r_s = 28.44~{\rm kpc}$~\cite{Einasto:1965czb,Navarro:2008kc}. On the
other hand, the cored profile put forward by Burkert, for which
$\beta = 0$, $\alpha =2$, and $r_s =12.67$, is motivated by
observations of galactic rotation
curves~\cite{Burkert:1995yz}. Profiles steeper than the NFW have also
been considered, e.g. the one by Moore and collaborators taking
$\alpha = 1.16$, $\beta = 1$, and
$r_s = 30.28~{\rm kpc}$~\cite{Moore:1999nt}. Herein, we take
$\rho_X = \rho_{\rm DM}$ and normalize to the local (solar) DM
density,
$\rho_X(r_\odot) = \rho^{{\rm DM}} _\odot = 0.3~{\rm GeV/cm}^3$, where
$r_\odot = 8.33~{\rm kpc}$ is the distance between the Earth and the
Galactic Center~\cite{Tanabashi:2018oca}. This leads to
$\rho_s/({\rm GeV\,cm^{-3}}) = 0.184, 0.033, 0.712, 0.105$ for the profiles
proposed by NFW, Einasto, Burket, Moore;
respectively~\cite{Cirelli:2010xx}. A comparison of these profiles is
given in Fig.~\ref{fig:1}. The ensuing discussion will be framed in
the context of NFW, and we will comment on the other profiles after
presenting our results.

\item The particle physics factor is built-in the fragmentation
  function of the SM particles produced by the $X$-decay. There is a
  general agreement among the various computational schemes (relying
  on either analytic approximations~\cite{Berezinsky:1998ed} or else
  Monte Carlo
  simulations~\cite{Birkel:1998nx,Berezinsky:2000up,Sarkar:2001se,Barbot:2002gt})
  proposed to describe the secondary spectra of SM particles produced
  via $X$-decay. Herein, we obtain the final state stable particle
  spectra by solving the DGLAP equations
  numerically~\cite{Gribov:rt,Gribov:ri,Dokshitzer:sg,Altarelli:1977zs}. As
  an illustration, in Fig.~\ref{fig:2} we show the resulting photon,
  proton, and neutrino ($\nu +\bar \nu$) spectra from $X \to q \bar q$
  decay. From
  the observational perspective, the salient features of the final
  state particles (photons, nucleons, and neutrinos) can be summarized
  as follows: {\it (i)}~the spectrum is flat
  ($dN/dE \propto E^{-1.9}$) and independent of the particle type,
  {\it (ii)}~the photon/nucleon ratio is $2\alt \gamma /N \alt 3$ and
  the neutrino/nucleon ratio is $3 \alt \nu/N \alt 4$; both of these
  ratios being quite independent of the energy.
\end{itemize}

The
expected energy distribution on Earth follows the
initial decay spectrum, whereas the angular distribution incorporates
the (uncertain) distribution of dark matter in the Galactic halo via
the line-of-sight integral~\cite{Dubovsky:1998pu, Evans:2001rv,Aloisio:2007bh,Kalashev:2017ijd}. The photon flux observed on Earth can be written as 
\begin{widetext}
\begin{equation}
J (E,\theta)=\frac{1}{4\pi} \ \frac{1}{\tau_X \ M_X} \ \frac{dN}{dE} \
\left \{ \, 
2 \ \int_{r_\odot \sin \theta}^{r_\odot} dr \ r  \
\frac{\rho_X(r)}{\sqrt{r^2 - r^2_\odot \, \sin^2\theta }} + 
 \int_{r_\odot}^{R_H} dr \ r \ \frac{\rho_X(r)}{\sqrt{r^2 - r^2_\odot
     \ \sin^2 \theta}} \right \}\,,
\end{equation}
\end{widetext}
where $\theta$ is the angle between the line of sight and the axis
defined by Earth and the Galactic center~\cite{Aloisio:2006yi}. Here, $\tau_X$ is the WIMPzilla
lifetime and $R_H = 260~{\rm kpc}$ is
the radius of the Galactic halo.

\begin{figure}[tb] 
    \postscript{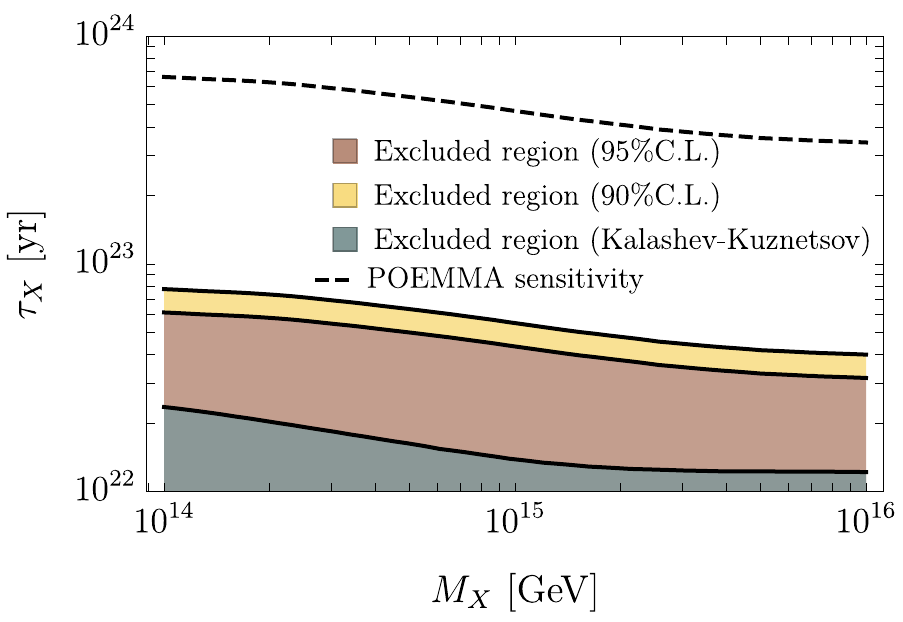}{0.99} 
\caption{Lower limit on the lifetime of SHDM particles together
  with the stereoscopic $\tau_X$ sensitivity (defined by the observation of one photon
  event above $10^{11.3}~{\rm GeV}$ in 5~yr of data collection) of
  POEMMA. The previous limit on $\tau_X$
  derived in~\cite{Kalashev:2016cre} is
  also shown for comparison.\label{figure}}
\end{figure}

Following~\cite{Kalashev:2016cre}, we normalize the flux integrating
over the whole sky ($0< \theta < \pi$) and averaging over the
directional exposure at the declination of the Auger
Observatory~\cite{Aab:2014ila}. For
$M_X = 1.7 \times 10^{16}~{\rm GeV}$ and
$\tau_X = 8.3 \times 10^{21}~{\rm yr}$, the intergral flux of photons
at the location of the Auger Observatory is
$J (>E_0) = 1.6 \times 10^{-4}~{\rm km^{-2}} \, {\rm yr^{-1}} \, {\rm
  sr^{-1}}$~\cite{Mikhail}. This is a factor of $1.75$ times smaller
than the integral flux of photons derived in~\cite{Aloisio:2015lva}
for the same value of $M_X$ and $\tau_X$, using $\alpha =3/2$,
$\beta =1$, and $r_s = 45~{\rm Mpc}$ as obtained
in~\cite{Berezinsky:1998rp}. Now, we compare the integral flux with
the upper limit derived in (\ref{gammabound}) to constrain the
$\tau_X-M_X$ parameter space.  Our results are encapsulated in
Fig.~\ref{figure}.  The growth of the final state stable particle
spectra with decreasing $x$ determines the functional form of
the constraint on $\tau_X$. For masses in the range,
$10^{14} 􏰹\alt M_X/{\rm GeV} \alt 10^{16}$, the lower limit (95\%C.L.)
on the lifetime of SHDM particles derived in this work, is a factor
$\agt 2$ more restrictive than previous
bounds~\cite{Kalashev:2016cre}; see
also~\cite{Gondolo:1991rn,Esmaili:2012us,Kuznetsov:2016fjt}. For
$M_X \alt 10^{14}~{\rm GeV}$, constraints on the diffuse photon flux
below $E_0$~\cite{Aab:2016agp,Abbasi:2018ywn} provide the most
restrictive bound on $\tau_X$~\cite{Kachelriess:2018rty}. A point
worth noting at this juncture is that the limit on $\tau_X$ is
completely independent of the $X$-production mechanism, and
consequently it applies to all SHDM models,
e.g.~\cite{Garny:2015sjg,Garny:2017kha}.

  There are a few  caveats to our calculation. On the one hand,
  it is important to emphasize that the limit derived in
  Fig.~\ref{figure} is calculated under the assumption that the
  photon-to-baryon relative exposure of the Auger surface detector
  array is equal to one.  This overly simplified assumption may
  overestimate the actual photon  exposure~\cite{Kalashev:2008dh,Aab:2015bza}. We defer a
  detailed description of the photon directional  exposure to the Auger
  Collaboration. On the other hand, it is important to note that the
  contribution from the nucleon flux to the all-particle intensity
  would  
  tend to compensate any possible reduction in the photon
  exposure. Indeed, we can derive a lower limit on $\tau_X$ using only
  the nucleon flux expected from the $X$-decay. A rough estimate of
  such a limit can be obtained through a re-scaling of the results
  shown in Fig.~\ref{figure}  by the $\gamma/N$ ratio. An additional compensation
  can be picked up by using also the Telescope Array (TA)
  observations. TA has accumulated an exposure
  $\sim 8,300~{\rm km^2 \, sr \, yr}$ without observation of events
  above $10^{11.3}~{\rm GeV}$~\cite{TheTelescopeArray:2018dje}. After removing
  the band of declination common to both experiments this becomes a
  $\sim 10\%$ effect.

\section{POEMMA discovery reach}
  
 In line with our stated plan, we now estimate the sensitivity of
  next generation UHECR experiments to detect signals of WIMPzillas.
  At present, the most advanced concept in pursuit of this objective
  is the Probe of Extreme Multi-Messenger Astrophysics
  (POEMMA)~\cite{Olinto:2017xbi}.  POEMMA will comprise two satellites
  flying in loose formation at 525 km altitudes, with stereoscopic
  UHECR observation mode and monocular Earth-limb viewing mode. In
  stereo fluorescence mode, the two detectors view a common immense
  atmospheric volume corresponding to approximately $10^{13}$ tons of
  atmosphere. The stereo mode yields roughly an order of magnitude
  increase in yearly UHECR exposure compared to that obtainable by
  ground observatory arrays and two orders of magnitude compared to
  ground fluorescence observations. In the limb-viewing mode, POEMMA
  reaches nearly $10^{10}$ gigatons. The stereoscopic sensitivity of
  POEMMA to probe the lifetime of SHDM is shown in Fig.~\ref{figure}. Detection of a
  extreme-energy photon would be momentous discovery.  If this were
  the case, POEMMA could be switched into limb-mode to rapidly
  increase statistics.

  It is also noteworthy that cosmic-ray showers initiated by
  extreme energy photons develop, on average, deeper in the atmosphere
  than air showers of the same primary energy initiated by
  protons~\cite{Anchordoqui:2018qom}. This is portrayed through the
  observable $X_{\rm max}$, which describes the atmospheric column
  depth at which the longitudinal development of a cosmic-ray shower
  reaches maximum. Of particular interest here, for energies $E\agt E_0$, the average $X_{\rm max}$
  of photon and proton showers differs by more than
  $100~{\rm g/cm^2}$~\cite{Risse:2007sd}. Ergo, while the expected
  monocular performance of POEMMA to identify the UHECR primary
  ($\Delta X_{\rm max} \sim 100~{\rm g/cm^2}$) is not as accurate as
  that for the stereo mode
  ($\Delta X_{\rm max} \lesssim 30~{\rm g/cm^2}$), it is still
  sufficient to characterize the $\gamma/N$ ratio.

  We now comment on the impact of the adopted DM profile in our
  calculations. Because we are averaging over the entire field of view
  of the experiments, the selection of the DM profile carries only a
  very small effect. This is visible in Fig.~\ref{fig:1} 
  where we show that the differences between the DM halo profiles are evident for
  angles $\theta \alt 10^\circ$. Indeed, the deviation from our
  results when considering the Burkert profile rather than the
  canonical NFW is
  about 10\%. Because the Galactic center is
well within the field of view of Auger the limit on $\tau_X$ is slightly
  relaxed when considering the Burket profile. One the other hand, the
  POEMMA sensitivity that averages over the orbital period is increased.

  \section{Conclusions}

  Thus far the various ongoing efforts to produce or detect WIMPs have
  not given us any promising clues, and moreover, as of today there
  have been no definitive hints for beyond SM physics at any
  accessible energy scale. This rather unexpected situation has
  motivated a new approach to understand the particle nature of DM.
  If the universe is fine-tuned then the natural mass range for the
  dark sector would be the Planck scale. Such SHDM can arise from
  String Theory or other high-energy phenomena, and the observed DM
  abundance  can be successfully produced during the
  inflationary epoch. We have studied the constraints on SHDM models
  given by recent UHECR observations. For masses in the
  range $10^{14} 􏰹\alt M_X/{\rm GeV} \alt 10^{16}$, we derived the
  strongest (95\% C.L.) limit on the lifetime of hadronically decaying
  SHDM particles. We also explored the prospects for WIMPzilla
  discovery with future observations of UHECRs. We end with an observation: in 5~yr of data
  collection POEMMA (in the limb-viewing mode) will have the potential
  to accumulate an unprecedented exposure
  ($\sim 10^6~{\rm km^2 \, sr \, yr}$) and become the ultimate
  WIMPzilla hunter.\\

  \acknowledgments{We are thankful to Mikhail Kuznetsov for insightful
    remarks on the manuscript and providing the 
    normalization for the photon flux expected to be observed at the
    declination of the Auger Observatory.  We also thank Sergey Troitsky
    as well as our colleagues of the Pierre Auger and POEMMA
    collaborations for some valuable discussions. This work has been
    supported by the by the U.S. National Science Foundation (NSF
    Grant PHY-1620661) and the National Aeronautics and Space
    Administration (NASA Grant 80NSSC18K0464).  Any opinions,
    findings, and conclusions or recommendations expressed in this
    material are those of the authors and do not necessarily reflect
    the views of the NSF or NASA.}

\end{document}